\documentclass[10pt, conference, compsocconf]{IEEEtran}
\usepackage{graphicx}

\usepackage{cite}
\usepackage{url}
\usepackage{xspace}

\usepackage[utf8]{inputenc}

\usepackage{mdframed}

\newcommand{\cran}{\emph{CRAN}\xspace}
\newcommand{\github}{\emph{GitHub}\xspace}
\newcommand{\git}{\emph{Git}\xspace}

\title{Anonymized e-mail interviews with
R package maintainers active on \cran and \github}
\author{Tom Mens\\
Software Engineering Lab, COMPLEXYS Research Institute\\
University of Mons\\
Avenue du Champ de Mars 6, 7000 Mons, Belgium}

\begin{document}
\maketitle

\begin{abstract}
This technical report accompanies the research article \cite{DecanEtAl2016} that empirically studies the problems related to inter-repository package dependencies in the R ecosystem of statistical computing, with a focus on R packages hosted on \cran and \github. That article extends our earlier research on the R package ecosystem, published in \cite{Claes2014-maintainer, Claes2014, DecanEtAl2015}.

The current report provides supplementary material, reproducing an anonymised and sanitised version of e-mail interviews that have been conducted in November 2015 with five active R package maintainers. The goal was to gain a better understanding in how R package maintainers develop and distribute their packages through \github and \cran.

All five interviewees were actively maintaining packages on \github, some were also active on \cran.
They have been selected based on their profile (the number of R packages they maintain on \github and/or \cran) as well as their gender (three interviewees were male, two were female).
\end{abstract}

%%%%%%%%%%%%%%%%%%%%%%%%%
\section{\color{blue}{Interview with first R package maintainer}}
\label{sec:first-interview}

At the time of the interview, this package maintainer had all his or her R packages only available on \github, and none of them on \cran.

\bigskip
{\color{blue}Why did you choose for \github as a development platform for your R
packages?}

Having been turned off by the inflexibilities and shortcomings of other
version control systems like SVN, TortoiseSVN and the like, I checked out
\git about 3-4 years ago and really came to love it. Then two things happened
that motivated moving over to \github:
\begin{enumerate}
\item I came across some work of [***ANONYMISED***], we started ``talking" and basically he
said: ``why don't you move your blog-post stuff over to \github, that way
everyone can contribute and has an easy to use single point of reference.
More and more people in the R universe are doing it, so come along".
\item The more I used \git, the more I felt I needed better support for
managing my TODO list of things I would like to do in my code. So I checked
out \github's issue mechanism and since then been totally hooked.
\end{enumerate}

\newpage
{\color{blue}Are there any specific reasons (technical or others) why your
packages are only made available through \github and not through \cran?}

These two aspects are probably the most compelling ones for me:
\begin{enumerate}
\item \github is \textbf{easy} to understand and use, \cran is not. I'd call myself a
more or less experienced R developer, still all those R CMD things and the
entire mechanism of publishing of \cran is somewhat intimidating for me. To
me \cran represents the ``old" way/style of doing stuff on the internet and in
coding, while \github, travisCI, CodeCov, StackOverflow etc. are \textbf{superb}
examples of how infrastructure technology can actually enable people to do
the things they actually want to do (produce awesome code/packages) instead
of spending weeks configuring things in order to get them to work. What adds
to this intimidation is the fact that you (or at least I) think ``oh man, my
package has to be \textbf{really} good in order to be on \cran; there are so many
expert packages out there, you're gonna be laughed at if it's not perfect".
That might just be me, but I would guess that other developers feel a
similar purely psychological or social barrier to using anything related to
\cran or the people behind it. I'd even argue that most of these perceived
barriers are causef by the general tone on the mailing list (some examples:
\url{http://bit.ly/1koPQax}, http://bit.ly/1MEum6l, http://bit.ly/1Wwmq6i).
Compare that to the tone on \url{http://stackoverflow.com/questions/tagged/r},
it's like day and night.

\item \github makes it so easy to reach out to other programmers for ideas,
feature requests, bugs in a \textbf{structured} yet very intuitive and fun way.
On \cran, you're bound to stick to e-mailing package authors and/or raising
attention through those tried-and-true, but IMO somewhat outdated mailing
lists in the R universe. Writing code in e-mails isn't fun, it's easy to
lose track of things, it's simply not the right mechanism of collaborating
on code and ideas involving code. 

\newpage
By structured I mean things like: 
\begin{itemize}
\item the issue tracking mechanism with all its people- and code-linking
capabilities and its flexible tagging system
\item  being able to fork an entire repo to play around with it and then issuing
a pull request to the author if you think you improved something
\item being able to reference specific "states of the world" (commits, releases
etc.)
\end{itemize}
\end{enumerate}

\smallskip
{\color{blue}Which of \github's software development facilities do you rely on (e.g.
version control, merging, continuous integration, issue tracking, and so
on)?} 

\begin{itemize}
\item version control
\item branching (I like this model very much:
\url{http://nvie.com/posts/a-successful-git-branching-model/}) 
\item continuous integration via travisCI
\item issue tracking
\item forking
\item pull requests
\end{itemize}

\smallskip
{\color{blue}Do you also use \github as a platform for distributing your R packages?}

Yes.

\smallskip
{\color{blue}Why and how?}

As I could never really bring up the courage and time to check out how
things actually need to work on a technical level in order to publish on
\cran, I simply use \github and having a lot of fun with it. Currently, almost
only through the use of \verb|devtools::install_github| as, for example, outlined
in [***ANONYMIZED***].
But recently I've also experimented with \textsf{drat} (\url{https://github.com/eddelbuettel/drat}) and that
looks promising. I'd probably always keep the \verb|devtools::install_github|
mechanism, but bring things more in line with R's repository philosophy by
using something like \textsf{drat}. I also plan to look into
\url{https://github.com/RevolutionAnalytics/checkpoint} and try to bring
everything together in a way that fits me.

\smallskip
{\color{blue}What are the \textbf{advantages} that you have perceived in using \github for
developing, distributing and installing R packages?}

\begin{itemize}
\item social interaction with coders 
\item conversations are ``spot-on" as it's easy to illustrate ideas with actual
code (including syntax highlighting, a \textbf{huge} enabler IMO) or link to actual
code
\item keeping track of things to do via the issue management system
\item quick and easy forking an re-integration via pull requests
\item in combination with \textsf{devtools}, things simply work, it's even fun
\end{itemize}

\newpage
{\color{blue}What are the \textbf{disadvantages} that you have perceived in using \github for
developing, distributing and installing R packages?}

Sometimes, it's a bit hard to mentally keep track of who's the originator
of a repository and who is just forking a repository. E.g., it always takes
me about 5 minutes to recall who is doing what with regard to
roxygen/roxygen2: if you search for ``roxygen github'' you end up here
\url{https://github.com/klutometis/roxygen}. If you search for "roxygen2 github",
you end up here \url{https://github.com/yihui/roxygen2}

It'd be nice to be able to sort repositories (manually or alphabetically)
and/or to tag one's own ``favorite repos". At times it seems that the main
user page is mainly designed for \textbf{others}, not for yourself (which is
okay, though). For example, [***ANONYMIZED***].

 I totally agree with Dirk's main point in
\url{http://dirk.eddelbuettel.com/code/drat.html} (Section "Drat Use Case 1:
GitHub"): \verb|devtools::install_github| is good, but we should stick to R's repo
logic. I'd be cool if something like this would already be supported
out-of-the box by \github.

I never had that experience, but I would imagine that as much as I'm
turned off by \cran, some people might be by having to use \github in order to
try out stuff I wrote.

\smallskip
{\color{blue}When developing R packages, have you encountered any specific problems
related to managing package dependencies?}

Yes. Whenever I write code, I entertain the idea that this code should be as
fit for productive usage as possible. Especially with respect to package
dependencies, the risk of things breaking at some point due to the fact that
a version of a dependency has changed without you knowing about it is
immense. That actually cost us weeks and months in a couple of professional
projects I was part of. While it's rather a philosophical than a technical
question how package dependencies are/should be handled in R, I personally
think it's \textbf{really} relevant to at least be \textbf{able} to be very specific and rigid
with regard to your dependencies. And I think the R universe could provide
better tools to fit the needs of developers and professionals out there in a
better way. But in that regard I like efforts such as
\url{https://github.com/rstudio/packrat} and
\url{https://github.com/RevolutionAnalytics/checkpoint} very very much. And I very
much like the fact that more and more people seem to be aware that this is
something one should manage systematically. Back in the days, I tried to
raise attention for it on R-devel mailing list without any success [***ANONYMIZED***].
I guess that was actually the point where I decided that I liked StackOverflow much better and would stop using the
mailing lists ;-))

A better systematic for dependency management together with making your
codebase more robust against changes in dependencies is the thing that I
actually spend most of my time cracking my brain about as I feel the R
universe could do a much better job in that respect. And I think it's still
one of the number one reasons why people are hesitant to use R in productive
rather than academic contexts while there is clearly a demand for using it
productively (e.g. \url{http://www.earl-conference.com/},
\url{http://www.r-bloggers.com/demo-r-in-sql-server-2016/}, etc.).

So I while there's clearly a purely technical aspect to the assessment of
R's dependency management system, I also feel that, on R's side, there is
great potential in learning from tried-and-true paradigms from the software
engineering world. A lot of R developers do not consider themselves
programmers as they usually picked up R at university studying something
other than software engineering/informatics etc. (as is also true for
myself: business administration and then applied statistics). So they
typically never heard of things like the SOLID principles of OOP, design
patterns, interfaces, base classes, class inheritance etc. While that's
totally fine, it does in fact really materialize when it comes to how well
things fit together in such a highly distributed system as R's add-on
package universe. And to that end, \github is also doing a superb job of
``educating people without them even knowing it". Or put differently, by
going out of the way of giving developers the opportunity to check out how
\textbf{other} developers are approaching certain tasks and ``how they are doing
things" people learn from each other in a natural/non-intrusive way. And
that's simply terrific! 

I personally learned \textbf{so much} by peeking into the \github projects of
\url{https://github.com/hadley}, \url{https://github.com/yihui},
\url{https://github.com/rstudio} (\url{https://github.com/jcheng5} and
\url{https://github.com/wch} in particular). They inspired my to do a better job
with regard to applying a better software design/architecture [***ANONYMIZED***].
Another milestone for my personal R dev skills was me being able to check out \textbf{how}
exactly Joe Cheng implemented that awesome reactivity stuff in
\url{http://shiny.rstudio.com/}  that he borrowed from \url{https://www.meteor.com/}. I
vividly remember watching his interview from useR! 2014
(\url{https://www.youtube.com/watch?v=uJm-its3ZWM}) and not being able to hold
back any longer with the desire to understand how this actually works. So I
checked out the \github repo of \textsf{shiny}, learned from it and created my own
little thing [***ANONYMIZED***]. While still being far from
a full-fledged package that I'd comfortably point someone to for actual use
in a project, it \textbf{definitely} brought my R game to higher level.

So there's this clear learning aspect that \github offers, as well. While
source code is technically also available on \cran, it takes much more effort
to check out the actual code of another programmer - so no one is doing it.
While there's arguably a higher code/package quality on \cran than on \github,
the platform/design keeps other developers from learning to become as good
as the top stars on \cran.

\newpage
%%%%%%%%%%%%%%%%%%%%%%%%%
\section{\color{blue}{Interview with second R package maintainer}}
\label{sec:second-interview}

At the time of the interview, this package maintainer had all his or her R packages only available on \github, and none of them on \cran.

\bigskip
{\color{blue}Why did you choose for \github for \textbf{developing} your R package(s)?}

We chose \github for development in order to take advantage of the version control/branches/etc software development tools offered.

\smallskip
{\color{blue}Do you also use \github as a platform for \textbf{distributing} your R packages?}

We are also using \github as a platform for distribution of the package. 
For development of packages, \github is very useful, and \cran is really not appropriate for development. For distribution, of course it is more of a pain for end users to have to get packages from anywhere other than \cran. 

\smallskip
{\color{blue}What are the \textbf{advantages} and \textbf{disadvantages} of using \github instead of \cran for developing, distributing and installing R packages?}

There are a few reasons we [***ANONYMIZED***] have only distributed very few of our packages on \cran:
\begin{enumerate}
\item \cran checks are not easy to pass. While we are in agreement with their requirements for vignettes and working examples, and we enforce these rules ourselves, there are many other small things that make it difficult to get a package posted to \cran. 
\item Many of our packages are quite specialized and have a potentially very small group of users. The work of getting onto \cran is not worth the reward in these cases. 
\item Every time you need to update the package on \cran you have to deal with all their hoops again. We believe in frequent updates, so this is time-consuming.
\item A package up on \cran must be maintained to a high standard. This is fine, but we have many packages with little to no maintenance budget. It is not worth it for us to have to deal with updates for each small revision of R if the package is still working fine for our core users. 
\end{enumerate}

That's the gist. I think that \cran is a great resource, and it is certainly great to have a central clearing house of packages that have been curated to some degree. The curation, however, is not without cost. I would note that Hadley Wickham (\textsf{ggplot}, etc) now has packages that he has chosen not to upload to \cran, for reasons similar to those I have cited above. 

\newpage
{\color{blue}When developing/maintaining R packages, have you encountered any specific problems related to managing package dependencies?}

Yes, it definitely is a problem sometimes. There have actually been a few times when I have rewritten a function in my own package because of that difficulty, especially with packages that themselves have many dependencies. Package creators, however, are getting more disciplined about only requiring packages that are essential, and using the ability to require a function rather than a package when possible. The biggest issue we have is multiple layers of dependencies, some of which are on \cran and some of which are not. 
That can be difficult to keep in sync, but usually, if your package is not on \cran, you can just keep it using the older dependency for a while until you have time to sort that issue.

\newpage
%%%%%%%%%%%%%%%%%%%%%%%%%
\section{\color{blue}{Interview with third R package maintainer}}
\label{sec:third-interview}

At the time of the interview, this package maintainer had all his or her R packages available on \cran, and several of these were also available on \github.

\bigskip
{\color{blue}Do you use Github as a platform for \textbf{developing} R package(s)?}

Yes. 

\smallskip
{\color{blue}Which of GitHub's software development facilities do you rely on (e.g. version
control, merging, continuous integration, issue tracking, and so on)?}

I've been using all of your examples, i.e. version control, branching, merging, Travis CI, issue tracking, also development in teams under an organizational account.

\smallskip
{\color{blue}What are the reasons (technical or others) why some of your packages
are only available through \github and not through \cran? }

I don't have any R packages that are on \github but not on \cran. I do have some on \cran that are not on \github but that's because I did not get to it yet. I plan to have all of my \cran packages in \github.

\smallskip
{\color{blue}Do you also use \github as a platform for \textbf{distributing} your R  packages? Why and how?}

I only do it if there is a an update of the package that users might need immediately, e.g. a bug fix. I just tell the users to get the updated version from \github and provide the link.

\smallskip
{\color{blue}For those of your packages that are available on \cran and \github,
is there a difference between both (e.g. stable version versus development version)?}

Yes. I usually have a ``cran" branch which matches the \cran version. A ``master" branch (or one that I set as default) would be the current development version. I might have some experimental branches as well.

\smallskip
{\color{blue}What are the \textbf{advantages} that you have perceived in using \github
for developing, distributing and installing R packages?}

\begin{itemize}
\item Version control.
\item Super easy branching (in comparison with e.g. Subversion).
[The two above are more about \git than \github.]
\item Having the code available from anywhere (since I work at different days on different computers).
\item Sharing the code with co-workers and users.
\item Travis CI allows to check the code against different versions of R, so that I don't need to re-install R every time a new development version is released.
\item Issue tracker helps me not to forget things that I wanted to fix or enhance.
\item I'm sure I'm forgetting something, but overall I really love \github!
\end{itemize}

\newpage
{\color{blue}What are the \textbf{disadvantages} that you have perceived in using Github for developing, distributing and installing R packages?}

I can't think of any that are specific to R packages. The only feature I was thinking would be nice is to be able to set specific branches as private/public, instead of having this setting to apply to the whole repository.

\smallskip
{\color{blue}When maintaining R packages, have you encountered any specific problems related to managing package dependencies?}

Nothing that would be related to \github.
I had one case where my package heavily depended on another package and after a while that package was removed from \cran and stopped being maintained. So I had to remove one of the main features of my package. Now I try to minimize dependencies on packages that are not maintained by ``established'' maintainers or by me ;-) Nowadays it has become a standard that PhD students in statistics develop an R package as a byproduct of their PhD thesis. But for students who leave academia or move to other projects it's often a low priority if at all to maintain their package.

\newpage
%%%%%%%%%%%%%%%%%%%%%%%%%
\section{\color{blue}{Interview with fourth R package maintainer}}
\label{sec:fourth-interview}

At the time of the interview, this package maintainer had all his or her R packages available on \github, and most of them (except for the experimental or immature ones) also on \cran.

\bigskip
{\color{blue}Which of \github's software development facilities do you rely on for \textbf{developing} R packages (e.g., collaborative software development, software versioning, branching, merging, continuous integration, issue tracking, bug tracking, developer communication, automated testing, and so on)?}

All of them.

\smallskip
{\color{blue}What are the reasons (technical or others) why some of your packages are only available through \github and not through \cran?}

Generally, all my packages go to \cran unless they are either experimental or still young.

\smallskip
{\color{blue}Do you also use \github as a platform for \textbf{distributing} your R  packages? Why and how?}

Yes, but just for development versions and experimental packages.
I tell people to use \textsf{devtools::install\_github().}

\smallskip
{\color{blue}For those of your packages that are available both on \cran and \github, is there a difference between both (e.g. stable version versus development version)?}

\cran = release\\
\github = development

\smallskip
{\color{blue}What are the \textbf{advantages} and \textbf{disadvantages}  that you have perceived in using \github for developing, distributing and installing R packages?}

\cran is used by far more people, but the process to get a package on to \cran is more onerous. Some of the
challenge is incidental (i.e. dealing with \cran maintainers can be
frustrating), but some of it is fundamental, as you must check that
your package will work in a wide variety of situations (and that it
doesn't break other packages without warning).

\github is lighter weight so it's easier to distribute development versions, but code
form \github also has less quality control.

\smallskip
{\color{blue}When maintaining R packages, have you encountered any specific problems related to managing package dependencies?}

It's a bit of a hassle when your package depends on other development
versions, but there are changes in the latest version of \textsf{devtools} to
make this easier.

\newpage
%%%%%%%%%%%%%%%%%%%%%%%%%
\section{\color{blue}{Interview with fifth R package maintainer}}
\label{sec:fifth-interview}

{\color{blue}Why did you choose for \github for \textbf{developing} your R packages?}

I like \github for easy collaboration with others on joint projects, and to make the history of my software and other projects easy to explore by others. For R packages, an additional advantage is easy installation via the \textsf{devtools} package.

\smallskip
{\color{blue}Do you also use \github as a platform for \textbf{distributing} your R packages?}

\github is mostly a complement to \cran. I do have packages that I've not put on \cran, just because I don't want to bother with the often tedious process of getting a package onto \cran and keeping the version on \cran up-to-date.

I have a number of  packages that are \github only, some because I'm no longer actively developing or using them [***ANONYMIZED***], and some because they're very early in development [***ANONYMIZED***].

\smallskip
{\color{blue}What are the \textbf{advantages} of using \github for \textbf{distributing} R packages?}

The main advantage to \github instead of \cran is that you can keep the package current without having to involve the \cran curators.

\smallskip
{\color{blue}What are the \textbf{disadvantages} of using \github instead of \cran for \textbf{distributing} R packages?}

The key disadvantages are exposure (it's harder for potential users to find R packages that are not on \cran) and need installation from source (\cran will pre-build package binaries). The latter is most important for packages that include C/C++ code that needs to be compiled. \cran also handles dependencies better.

\smallskip
{\color{blue}What are the main \textbf{problems} you have experienced (or are currently experiencing) with the use of \textbf{\github} (in the context of R packages)?}

Regarding the use of \github, the main issue is specifying the set of tools and dependent packages that users need to install before installing the package, particularly in the case of compiled code.

\smallskip
{\color{blue}What are the main \textbf{problems} you have experienced (or are currently experiencing) with the use of \textbf{\cran} (in the context of R packages)?}

Regarding \cran, the curators have very specific rules about the specification of authors and of licenses that I don't entirely agree with. And one of the curators can be rather mean. Because of the heavy human involvement, one needs to be careful about not updating the \cran version of packages too often.

\newpage
{\color{blue}What are the main \textbf{advantages} you have experienced (or are currently experiencing) with the use of \textbf{\github} (in the context of R packages)?}

The main advantages of \github are in collaboration with others (receiving and incorporating pull requests, or issues), and of keeping the package current.

\smallskip
{\color{blue}What are the main \textbf{advantages} you have experienced (or are currently experiencing) with the use of \textbf{\cran} (in the context of R packages)?}

The main advantage of \cran is in ease of installation by users, particularly of packages with compiled code.

\bigskip

%\bibliographystyle{IEEEtran}
%\bibliography{biblio}  % sigproc.bib is the name of the Bibliography in this case

\begin{thebibliography}{1}
\providecommand{\url}[1]{#1}
\csname url@samestyle\endcsname
\providecommand{\newblock}{\relax}
\providecommand{\bibinfo}[2]{#2}
\providecommand{\BIBentrySTDinterwordspacing}{\spaceskip=0pt\relax}
\providecommand{\BIBentryALTinterwordstretchfactor}{4}
\providecommand{\BIBentryALTinterwordspacing}{\spaceskip=\fontdimen2\font plus
\BIBentryALTinterwordstretchfactor\fontdimen3\font minus
  \fontdimen4\font\relax}
\providecommand{\BIBforeignlanguage}[2]{{%
\expandafter\ifx\csname l@#1\endcsname\relax
\typeout{** WARNING: IEEEtran.bst: No hyphenation pattern has been}%
\typeout{** loaded for the language `#1'. Using the pattern for}%
\typeout{** the default language instead.}%
\else
\language=\csname l@#1\endcsname
\fi
#2}}
\providecommand{\BIBdecl}{\relax}
\BIBdecl

\bibitem{DecanEtAl2016}
\BIBentryALTinterwordspacing
A.~Decan, T.~Mens, M.~Claes, and P.~Grosjean, ``When {GitHub} meets {CRAN}: An
  analysis of inter-repository package dependency problems,'' in \emph{Int'l
  Conf. Software Analysis, Evolution, and Reengineering (SANER 2016)},
  vol.~1.\hskip 1em plus 0.5em minus 0.4em\relax {IEEE}, March 2016, pp.
  493--504. \url{http://dx.doi.org/10.1109/SANER.2016.12}
\BIBentrySTDinterwordspacing

\bibitem{Claes2014-maintainer}
\BIBentryALTinterwordspacing
M.~Claes, T.~Mens, and P.~Grosjean, ``{maintaineR}: A web-based dashboard for
  maintainers of {CRAN} packages,'' in \emph{Int'l Conf. Software Maintenance
  and Evolution ({ICSME} 2014)}, Sep. 2014, pp. 597--600.
  \url{http://dx.doi.org/10.1109/ICSME.2014.104}
\BIBentrySTDinterwordspacing

\bibitem{Claes2014}
\BIBentryALTinterwordspacing
------, ``On the maintainability of {CRAN} packages,'' in \emph{Int'l Conf.
  Software Maintenance, Reengineering and Reverse Engineering ({CSMR-WCRE}
  2014)}, Feb. 2014, pp. 308--312.
  \url{http://dx.doi.org/10.1109/CSMR-WCRE.2014.6747183}
\BIBentrySTDinterwordspacing

\bibitem{DecanEtAl2015}
\BIBentryALTinterwordspacing
A.~Decan, T.~Mens, M.~Claes, and P.~Grosjean, ``On the development and
  distribution of {R} packages: An empirical analysis of the {R} ecosystem,''
  in \emph{Proceedings of the 2015 European Conference on Software Architecture
  Workshops}, ser. ECSAW '15.\hskip 1em plus 0.5em minus 0.4em\relax New York,
  NY, USA: ACM, 2015, pp. 41:1--41:6.
  \url{http://doi.acm.org/10.1145/2797433.2797476}
\BIBentrySTDinterwordspacing

\end{thebibliography}
\providecommand{\noopsort}[1]{}

\end{document}